\documentclass[showpacs,showkeys]{revtex4}
\usepackage{graphicx}

\usepackage{amsmath}
\usepackage{amssymb}
\usepackage{array}
\usepackage{color}

\begin{document}

\title{Gluon field distribution between three infinitely spaced quarks}

\author{Vladimir Dzhunushaliev}
\affiliation{Institute for Basic Research,
Eurasian National University,
Astana, 010008, Kazakhstan \\
and \\
Institute of Physics of National Academy of Science
Kyrgyz Republic, 265 a, Chui Street, Bishkek, 720071,  Kyrgyz Republic}
\email[Email: ]{vdzhunus@krsu.edu.kg}

\date{\today}

\begin{abstract}
The gluon field distribution between three infinitely spaced quarks is obtained. The field distribution between every pair is approximated by an infinite flux tube filled with a color longitudinal electric field. The origin of the color electric field is a nonlinear term in the field strength of nonabelian gauge field. Such construction can be considered as a rough model of three quarks in a nucleon.
\end{abstract}


\maketitle

\section{Introduction}

One interesting problem in quantum chromodynamics is the calculation of gluon distribution between quarks. If we calculate the field distribution between two quarks we will obtain a hypothesized flux tube that is the main point of the confinement problem in quantum chromodynamics. The problem is analogous to the calculation of a static electric field between electron and positron in electrodynamics with one essential difference. The field distribution between quarks and between electric charges are essentially different: the gluon field between two quarks is confined into a tube but the electric field is distributed in the whole space. If we will calculate the field distribution between three quarks we will obtain a gluon distribution between quarks in a hadron. In electrodynamics we have not any analogous problem because quarks may have three different kind of charge: \textcolor{red}{red}, \textcolor{green}{green} and \textcolor{blue}{blue}. But the electric charge has not a color index.

Here we will calculate the gluon field distribution between three infinitely separated quarks. But why infinitely separated ? The answer is that we have the field distribution between two quarks \cite{Dzhunushaliev:2010qs}. In this case in a first approximation we can obtain the field distribution between three quarks as three flux tubes connecting three infinitely spaced quarks.

The analytical calculations of the gluon field distribution between quarks can not be carried out in the consequence of missing of a nonperturbative calculation technique in quantum field theory. There exist calculations of the field distribution between quarks on lattice, see for example, Ref. \cite{Bornyakov:2004hm}. The lattice calculations allow us to investigate a gauge field distribution between quarks. But the problem in lattice calculations is that we have not any analytical expression for investigated quantities such as: fields; potentials; energy, action, monopole densities and so on. Another problem is that we are not self-confident yet that there exists a limit when a grid size of lattice tends zero.

Our calculations are following: (a) we will recall how one can obtain the flux tube following to Ref. \cite{Dzhunushaliev:2010qs}, (b) we will clarify what is the origin of a color electric field in the flux tube and (c) we will connect three infinitely spaces quarks with these flux tubes filled with color electric fields.

\section{Flux tube}

Nonperturbative quantization is unresolved problem in modern theoretical physics. Here for the calculations in quantum chromodynamics we use following approximate method \footnote{In fact this method was proposed by Heisenberg as a nonperturbative quantization method in Ref. \cite{heisenberg}}: we calculate SU(3) gluon condensate
$\left\langle \hat {\mathcal F}^{B\mu\nu} \hat {\mathcal F}^{B}_{\mu\nu} \right\rangle$ which up to a sign is the quantum averaged SU(3) Lagrangian
\begin{equation}
	\left\langle \mathcal L_{SU(3)} \right\rangle = - \frac{1}{4 g^2}
	\left\langle
		\hat {\mathcal F}^{B\mu\nu} \hat {\mathcal F}^{B}_{\mu\nu}
	\right\rangle
\label{1-10}
\end{equation}
where $\mathcal F^B_{\mu \nu} = \partial_\mu A^B_\nu - \partial_\nu A^B_\mu + g f^{BCD} A^C_\mu A^D_\nu$
is the field strength; $B,C,D = 1, \ldots ,8$ are the SU(3) color indices; $g$ is the coupling constant; $f^{BCD}$ are the structure constants for the SU(3) gauge group and $g$ is the coupling constant. In Ref. \cite{Dzhunushaliev:2010qs} it is shown that using some assumptions and approximations one can reduce SU(3) Lagrangian \eqref{1-10} to an effective Lagrangian
\begin{equation}
	\mathcal L_{eff} =
	- \frac{1}{4 g^2} F^{a}_{\mu\nu} F^{a \mu\nu} +
	\frac{1}{2} \left| \phi_\mu \right|^2 -
	\frac{\lambda}{4} \left(
		\left| \phi \right|^2 - \phi_\infty^2
	\right)^2 +
	\frac{1}{2} A^a_\mu A^{a \mu}  \left| \phi \right|^2 -
	\frac{1}{2} m^2_a A^a_\nu A^{a \mu}
\label{1-20}
\end{equation}
where $\phi$ is a complex scalar field describing quantum fluctuations of  $A^m_\mu \in SU(3)/SU(2)$ gauge field components; $\phi_\infty$ and $m_a$ are some parameters.

In order to obtain effective Lagrangian \eqref{1-20} we have used following assumptions:
\begin{itemize}
  \item The first assumption is that the SU(3) gauge potential $A^B_\mu \in SU(3), B=1,2, \cdots , 8$ can be separated on two parts:
\begin{itemize}
	\item the first one is the gauge components $A^b_\mu \in SU(2) \subset SU(3)$ which is in a classical state;
	\item the second one is $A^m_\mu \in SU(3)/SU(2)$ and it is in a quantum state.
\end{itemize}
One can say that $A^b_\mu$ is in an ordered phase and $A^m_\mu$ is in a disordered phase.
  \item The second assumption is that the 2-point Green's function of color quantum field can be approximately presented as the product of scalar fields with some coefficients having color and Lorentzian indices.
  \item The third assumption is that the 4-point Green's function can be decomposed as the product of two 2-point Green's functions.
\end{itemize}
The second and third assumptions gives rise to the fact that 2 and 4-point Green's functions are 
\begin{eqnarray}
	\left( G_2 \right)^{mn}_{\mu \nu}(x_1, x_2) &=&
	\left\langle
		A^m_\mu (x_1) A^n_\nu (x_2)
	\right\rangle \approx C^{mn}_{\mu \nu} \phi(x_1) \phi^*(x_2) +
	\tilde m^{mn}_{\mu \nu},
\label{1-24}\\
    G^{mnpq}_{\mu \nu \rho \sigma}(x_1, x_2, x_3, x_4) &=&
	\left\langle
		A^m_\mu (x_1) A^n_\nu (x_2) A^p_\rho (x_3) A^q_\sigma (x_4)
	\right\rangle \approx
	\left( G_2 \right)^{mn}_{\mu \nu}(x_1, x_2)
    \left( G_2 \right)^{pq}_{\rho \sigma}(x_1, x_2).
\label{1-28}
\end{eqnarray}
The field equations for the Lagrangian \eqref{1-20} are
\begin{eqnarray}
  \frac{1}{4g^2}D_\nu F^{a\mu\nu} &=& \left(
  	\left| \phi \right|^2 - m^2_a
  \right) A^{a \mu}, \text{ no summation over } a ,
\label{1-30}\\
  \phi_\mu^{; \mu} &=& -\lambda \phi
  \left( \left| \phi \right|^2 - \phi_\infty^2
  \right) + A^a_\mu A^{a \mu} \phi .
\label{1-40}
\end{eqnarray}
In Ref. \cite{Dzhunushaliev:2010qs} the index $a=1,2,3$, i.e. the SU(2) subgroup on generators $\lambda_{1,2,3}$ is spanned ($\lambda_B$ are Gell-Mann matrices that are SU(3) generators). Of coarse it is not necessary. It is necessary only that the index $a$ will be the index of some SU(2) subgroup. For example, it can be:
$a=(1,2,3); (1,4,7); (1,5,6); (2,4,6); (2,5,7); (3,4,5); (3,6,7)$. It follows form the fact there are following SU(3) structure constants
\begin{equation}
    f_{123} = 1, \quad f_{147} = - f_{156} = f_{246} = f_{257} =
    f_{345} = - f_{367} = \frac{1}{2}
\label{1-55}
\end{equation}
which describe SU(2) subgroups $SU(2) \subset SU(3)$. This difference is unessential because we can redefine coupling constant $g$ in such a way that to kill the factor $f_{abc} = 1/2$.

If we choose SU(2) subgroup spanned on SU(3) generators $\lambda_{1,4,7}$ then the solution one can search in the following form
\begin{equation}
    A^1_t(\rho) = f(\rho) ; \quad A^4_z(\rho) = v(\rho) ;
    \quad \phi(\rho) = \phi(\rho)
\label{1-50}
\end{equation}
here $z, \rho , \varphi$ are cylindrical coordinate system. The substitution
into equations \eqref{1-30} \eqref{1-40} gives us
\begin{eqnarray}
    f'' + \frac{f'}{x} &=& f \left( \phi^2 + v^2 - m^2_1 \right),
\label{1-60}\\
    v'' + \frac{v'}{x} &=& v \left( \phi^2 - f^2 - m^2_2 \right),
\label{1-70}\\
    \phi'' + \frac{\phi'}{x} &=& \phi \left[ - f^2 + v^2
    + \lambda \left( \phi^2 - \mu^2 \right)\right]
\label{1-80}
\end{eqnarray}
here we redefined $g \phi /\phi(0) \rightarrow \phi$, $f /\phi(0)  \rightarrow f$,
$v /\phi(0)  \rightarrow v$, $g \phi_\infty /\phi(0) \rightarrow \mu$,
$g m_{1,2} /\phi(0)  \rightarrow m_{1,2}$, $\rho \sqrt{\phi(0)}  \rightarrow x$. The color electric and magnetic fields are
\begin{equation}
  F^1_{t \rho} = -f', \quad F^4_{z \rho} = - v', \quad
  F^7_{tz} = fv .
\label{1-90}
\end{equation}
The most important here is that the color longitudinal electric field $F^7_{tz}$ is not a gradient of a gauge potential but is the consequence of a nonlinear term in the definition of the field tensor
$F^a_{\mu \nu}$.

\begin{figure}[h]
\begin{minipage}[t]{.45\linewidth}
  \begin{center}
  \fbox{
  \includegraphics[width=.85\linewidth]{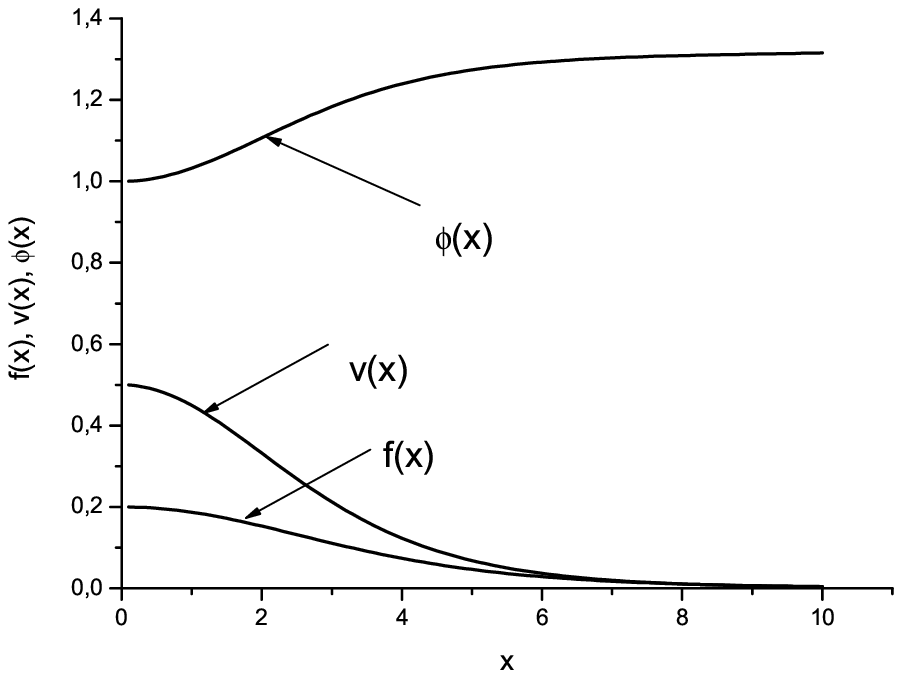}}
  \caption{The profiles of functions $f^*(x), v^*(x),\phi^*(x)$}
  \label{fig1}
  \end{center}
\end{minipage}\hfill
\begin{minipage}[t]{.45\linewidth}
  \begin{center}
  \fbox{
  \includegraphics[width=.85\linewidth]{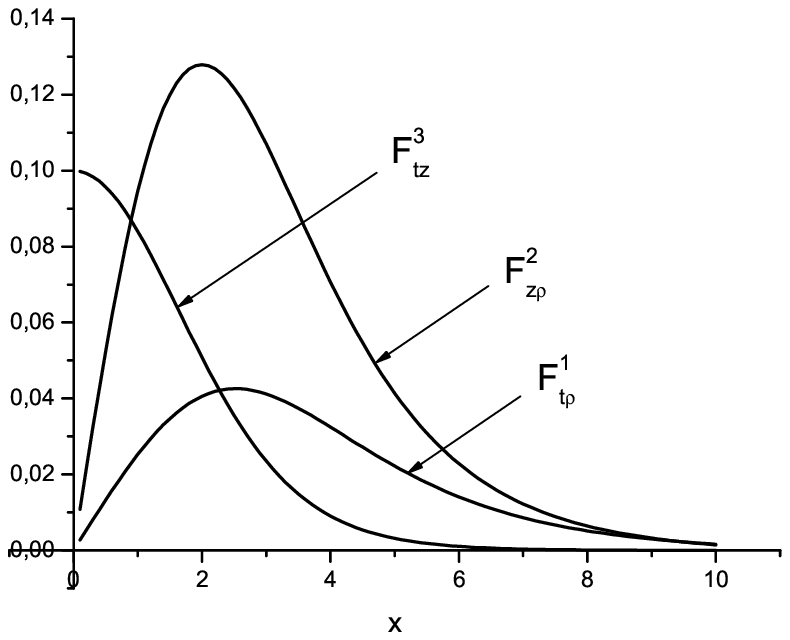}}
  \caption{The profiles of color fields
  $F^1_{t\rho}(x) = -f', F^4_{z\rho}(x) = -v', F^7_{tz}(x) = f(x) v(x)$.}
  \label{fig2}
  \end{center}
\end{minipage}\hfill
\end{figure}

For the numerical calculations we choose the following parameters values
\begin{equation}
	\lambda = 0.1, \quad
	\phi (0) = 1, \quad
	v(0) = 0.5, \quad
	f(0) = 0.2.
\label{1-100}
\end{equation}
The numerical values of these parameters are not unique but qualitative behavior of the functions $f(r),v(r),\phi(r)$ are the same. The solution of Eq's \eqref{1-60}-\eqref{1-80} with parameters \eqref{1-100} is presented in Fig's \ref{fig1} and \ref{fig2}

\section{The sources of longitudinal electric field at the end of the flux tube}
\label{tube}

Now we would like to discuss what is the source of the above mentioned longitudinal electric field $F^7_{tz}$. Firstly we would like to calculate the flux of the chromoelectric field $F^7_{tz}$
\begin{equation}
	\Phi = 2 \pi \int \limits_0^\infty \rho F^3_{tz} d \rho \neq 0.
\label{2-05}
\end{equation}
it is nonzero and consequently at the ends of the color flux tube should be color sources of this color field. For the determination of the color sources we will remind the SU(3) Lagrangian with quarks
\begin{equation}
	\mathcal L_{QCD} = - \frac{1}{2}
	\mathcal F^{B\mu\nu} \mathcal F^{B}_{\mu\nu} +
    \bar \psi \left(
        p_\mu + \frac{g}{2} A^B_\mu \lambda^B
    \right) \psi
\label{2-10}
\end{equation}
here we follow to \cite{greiner}; $\lambda^B/2$ are the SU(3) generators; spinor $\psi$ describe quarks \begin{equation}
	\psi = \begin{pmatrix}
        \textcolor{red}{\psi_r} \\
        \textcolor{green}{\psi_g} \\
        \textcolor{blue}{\psi_b}
    \end{pmatrix}
\label{2-20}
\end{equation}
where spinors $\psi_{\textcolor{red}{r},\textcolor{green}{g},\textcolor{blue}{b}}$ describe \textcolor{red}{red}, \textcolor{green}{green} and \textcolor{blue}{blue} quarks correspondingly. The term $\bar \psi \lambda^B \psi$ describes the source for the $F^B_{\mu \nu}$ field.

In our case we have $F^7_{tz}(x)$. Consequently the source of the color nonabelian longitudinal electric field has the form
\begin{equation}
	\bar \psi \lambda^7 \psi = \left(
        \textcolor{red}{\bar \psi_r}, \textcolor{green}{\bar \psi_g}, \textcolor{blue}{\bar \psi_b}
    \right)
    \begin{pmatrix}
        0   &   0   &   0   \\
        0   &   0   &   -i  \\
        0   &   i   &   0   \\
    \end{pmatrix}
    \begin{pmatrix}
        \textcolor{red}{\psi_r} \\
        \textcolor{green}{\psi_g} \\
        \textcolor{blue}{\psi_b}
    \end{pmatrix} =
    i \left(
        \textcolor{blue}{\bar \psi_b} \textcolor{green}{\psi_g} -
        \textcolor{green}{\bar \psi_g} \textcolor{blue}{\psi_b}
    \right).
\label{2-30}
\end{equation}
Thus at the ends of this flux tube there are quarks
$\left(\textcolor{red}{\psi_r}, \textcolor{green}{\psi_g}, \textcolor{blue}{\psi_b} \right)^T$ and
$\textcolor{green}{\psi_g}, \textcolor{blue}{\psi_b}$ components generate the \textcolor{green}{green}/\textcolor{blue}{blue} color longitudinal electric field $F^7_{tz}(x)$.

\section{Flux tubes between three quarks spaced at the infinity}

\begin{figure}[h]
  \begin{center}
  \fbox{
  \includegraphics[width=.5\linewidth]{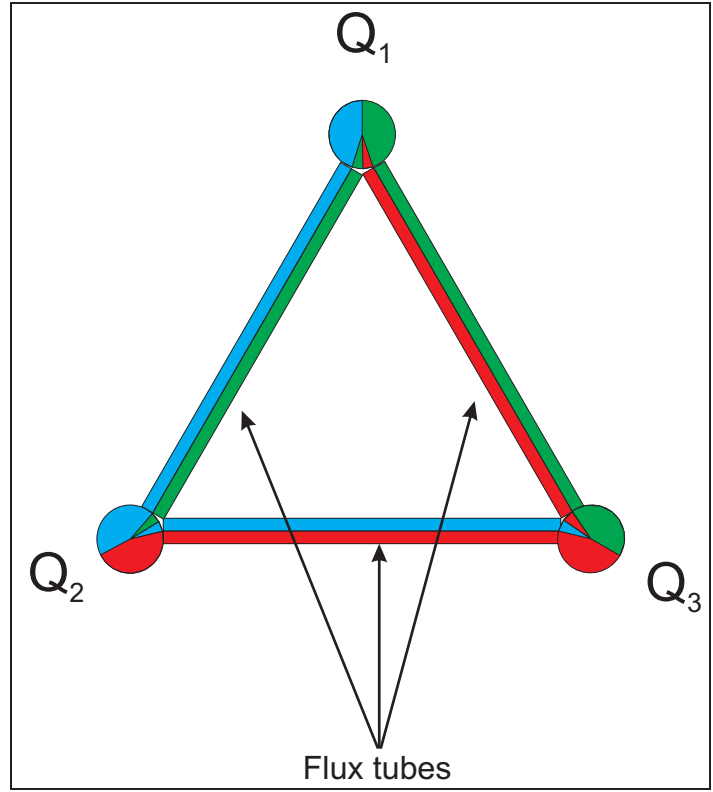}}
  \caption{Three infinitely spaced quarks connected with color flux tubes.}
  \label{fig3}
  \end{center}
\end{figure}

Now is the time to discuss the simplest model of gluon field distribution between three quarks. We simplify such model moving off the quarks to the infinity. In this case we can use the flux tube obtained in the section \ref{tube}. Let us consider three quarks $Q_{1,2,3}$ in Fig. \ref{fig3}.
Let us assume that the flux tube with $F^7_{tz}(x)$ longitudinal electric field is stretched between $Q_1$ and $Q_2$ quarks.

Now we change SU(2) subgroup spanned on $\lambda_{1,4,7}$ on the SU(2) subgroup spanned on $\lambda_{1,5,6}$. Then one can use ansatz
\begin{equation}
    A^1_t(\rho) = f(\rho) ; \quad A^6_z(\rho) = v(\rho) ;
    \quad \phi(\rho) = \phi(\rho).
\label{4-10}
\end{equation}
For such choice the color electric and magnetic fields will be
\begin{equation}
  F^1_{t \rho} = -f', \quad F^6_{z \rho} = - v', \quad
  F^5_{tz} = fv .
\label{4-20}
\end{equation}
Consequently the source of the color longitudinal electric has the form
\begin{equation}
	\bar \psi \lambda^5 \psi = \left(
        \textcolor{red}{\bar \psi_r}, \textcolor{green}{\bar \psi_g}, \textcolor{blue}{\bar \psi_b}
    \right)
    \begin{pmatrix}
        0   &   0   &   -i   \\
        0   &   0   &   0   \\
        i   &   0   &   0   \\
    \end{pmatrix}
    \begin{pmatrix}
        \textcolor{red}{\psi_r} \\
        \textcolor{green}{\psi_g} \\
        \textcolor{blue}{\psi_b}
    \end{pmatrix} =
    i \left(
        \textcolor{blue}{\bar \psi_b} \textcolor{red}{\psi_r} -
        \textcolor{red}{\bar \psi_r} \textcolor{blue}{\psi_b}
    \right).
\label{4-30}
\end{equation}
Thus at the ends of this flux tube there are quarks
$Q_{2,3} = \left(\textcolor{red}{\psi_r}, \textcolor{green}{\psi_g}, \textcolor{blue}{\psi_b} \right)^T$ and their
$\textcolor{red}{\psi_r}, \textcolor{blue}{\psi_b}$ components generate the \textcolor{red}{red}/\textcolor{blue}{blue} color longitudinal electric field $F^5_{tz}(x)$.

In the same way one can choose the SU(2) subgroup spanned on SU(3) generators $\lambda_{4,6,2}$. Then
\begin{equation}
    A^4_t(\rho) = f(\rho) ; \quad A^6_z(\rho) = v(\rho) ;
    \quad \phi(\rho) = \phi(\rho).
\label{4-60}
\end{equation}
For such choice the color electric and magnetic fields will be
\begin{equation}
  F^4_{t \rho} = -f', \quad F^6_{z \rho} = - v', \quad
  F^2_{tz} = fv .
\label{4-70}
\end{equation}
Consequently the source of the color longitudinal electric has the form
\begin{equation}
	\bar \psi \lambda^2 \psi = \left(
        \textcolor{red}{\bar \psi_r}, \textcolor{green}{\bar \psi_g}, \textcolor{blue}{\bar \psi_b}
    \right)
    \begin{pmatrix}
        0   &   -i   &  0   \\
        i   &   0   &   0   \\
        0   &   0   &   0   \\
    \end{pmatrix}
    \begin{pmatrix}
        \textcolor{red}{\psi_r} \\
        \textcolor{green}{\psi_g} \\
        \textcolor{blue}{\psi_b}
    \end{pmatrix} =
    i \left(
        \textcolor{green}{\bar \psi_g} \textcolor{red}{\psi_r} -
        \textcolor{red}{\bar \psi_r} \textcolor{green}{\psi_g}
    \right).
\label{4-80}
\end{equation}
Thus at the ends of this flux tube there are quarks
$Q_{1,3} = \left(\textcolor{red}{\psi_r}, \textcolor{green}{\psi_g}, \textcolor{blue}{\psi_b} \right)^T$ and their
$\textcolor{red}{\psi_r}, \textcolor{green}{\psi_g}$ components generate the \textcolor{red}{red}/\textcolor{green}{green} color longitudinal electric field $F^5_{tz}(x)$.

Finally, connecting three quarks $Q_{1,2,3}$ with corresponding color flux tubes we will obtain the construction from three quarks presented in Fig. \ref{fig3}. 

\section{Conclusions and discussion}

Thus we have obtained the field distribution of the SU(3) gauge field between three infinitely spaced quarks, see Fig. \ref{fig3}. In order to obtain such distribution we have used the flux tube solution obtained by a nonperturbative quantization method \cite{heisenberg}.

The problem of obtaining the field distribution between three color charges has not any analogy to a similar problem in electrodynamics. The point is that the color charges have additional degrees of freedom: color indices and this fact gives rise to principal difference between nonabelian and abelian gauge fields. The difference can be illustrated for the field distribution for three SU(3) color charges - \textcolor{red}{r}, \textcolor{green}{g}, \textcolor{blue}{b} quarks. Let us consider Fig. \ref{fig3}. For every pair $Q_i, Q_j$ the gluon field is concentrated in a tube (in the first approximation). A color longitudinal electric field is created from the pair either \textcolor{red}{r}, \textcolor{green}{g} or \textcolor{green}{g}, \textcolor{blue}{b} or \textcolor{red}{r}, \textcolor{blue}{b} quark degrees of freedom. It allows us to construct field distribution between three quarks similar to the Fig. \ref{fig3}.

The construction of the field distribution of a nonabelian gauge fields between three quarks presented here is approximate one. More exact description one can obtain considering the quarks spaced at a finite distance. Such construction can be considered as a model of a nucleon filled with three quarks. The model considered here is a rough approximation for nucleon.

\section*{Acknowledgements}

I am grateful to the Research Group Linkage Programme of the Alexander von Humboldt Foundation for the support of this research.

\end{document}